\documentclass[a4paper,fleqn,usenatbib]{mnras}

\usepackage[T1]{fontenc}
\usepackage{ae,aecompl}

\usepackage{graphicx}	
\usepackage{amsmath}	
\usepackage{amssymb}	

\title[The lonely Moon and double asteroids]{The lonely Moon, double asteroids, and multiple collisions}

\author[N.N. Gorkavyi$^{1}$ and T. A. Taidakova$^{2}$]{
N.N. Gorkavyi$^{1}$\thanks{E-mail: nick.gorkavyi@gmail.com}
and T. A. Taidakova$^{2}$
\\
$^{1}$Science Systems and Applications, Lanham, 20706 USA
\\
$^{2}$CCS LLC, Haymarket, VA, 20169 USA
}

\pubyear{2021}

\begin{document}
\label{firstpage}
\pagerange{\pageref{firstpage}--\pageref{lastpage}}
\maketitle

\begin{abstract} The Chelyabinsk meteor (02/15/2013) sailed over skies in a streak of light 
that was momentarily brighter than the Sun. 
The book \textbf {Chelyabinsk Superbolide} (Eds. Nick Gorkavyi, Alexander Dudorov, Sergey Taskaev, 
Springer Praxis Books, 2019) chronicles Chelyabinsk’s tale of recovery and discovery from 
the minds of many of the scientists who studied the superbolide, leading field experiments 
and collecting meteorites and meteorite dust across the region. The Chelyabinsk superbolide 
is a complex and multi-aspect phenomenon. 
The Appendix to the book examines the role of cosmic collisions in the formation and 
evolution of such bodies of the solar system as the Moon, double asteroids and irregular 
satellites of giant planets.
\end{abstract}




Space collisions, of which the fall of the Chelyabinsk bolide to Earth is an example, 
are a catalyst of planetary evolution. Without considering the process of multiple  
collisions,  it  is  impossible  to  understand  the  key  issues  of  celestial 
arrangement, in other words, the origin of asteroids and the natural satellites of 
planets, including the Moon. How did the Moon appear? This issue has concerned astronomy for over 2,500 
years, and begins with the question:

\textbf{Where did the matter, from which the Moon formed, come from?}

A related question has not been puzzled over for quite as long, a mere 200 
years:

\textbf{How did the belt of asteroids appear between Mars and Jupiter and why did the belt 
of asteroids not shape into a planet?}

It has been found that the asteroid belt did not accumulate into a planet because 
its current weight is only equal to the weight of our Moon, or 0.05\% of 
Earth’s mass. Therefore, the question about the origin of the asteroid belt can be 
reformulated: 

\textbf{Where did the 99.9\% of the original matter of the asteroid belt go to?}

The problem of the asteroid matter seems to be opposite to that of the source of 
matter for the formation of the Moon, but in reality, these are two closely related 
questions. The significant moment in this confusing story was the discovery of 
double asteroids, or asteroids with natural satellites, i.e. moons. From this, a new question arose: 

\textbf{How did the moons of asteroids, or double asteroids, appear?}

This question shed some unexpected light 
on the problem of the formation of the Moon and the asteroid belt.
However, let us return to square one. Until 1610, the Moon was believed to be 
unique; there were no other natural satellites known in the Solar System. Then, 
400 years ago, Galileo Galilei looked into his telescope and observed four new 
moons  near  Jupiter,  beginning  a  new  age  of  research  into  the  Solar  System. 
The Moon was no long a unique phenomenon, but in spite of the rapid growth in 
the number of discovered moons, the Moon did keep some of its exclusivity: all 
the new moons discovered were only found near the gas giants Jupiter, Saturn, 
Uranus, and Neptune, with the exception of the two tiny satellites of Mars, Phobos 
and Deimos. They were the only rivals to our Moon as satellites of planets with a 
solid surface and were only discovered relatively recently, in 1877. In terms of the 
relative mass (mass of the satellite divided by the mass of the planet), our Moon 
surpassed all the other known moons of the Solar System.

Its reputation as a unique natural satellite, different from any other, brought 
about a unique origin model for the Moon: the theory of a mega-impact. The theory 
was proposed by the Canadian geologist Reginald Daly in 1946 and was later 
was given a second life in a 1975 article by American scientists Bill Hartmann and 
Donald Davis. According to this theory (see its present-day discussion in \citet{Canup}), 
the huge planet Theia collided with Earth 4.5 billion 
years ago. Theia was a similar size to Mars (diameter about 7,000 km) and as a 
result of this truly unique event, some splinters from the collision went into an 
orbit around Earth, or more accurately, into an orbit as the new body resulting 
from the merging of Earth and Theia.

Although  this  model  became  almost  universally  accepted,  geochemists 
J.H. Jones and H. Palme \citep{Jones} opposed the mega-impact theory \citep{Canup2}. 
The geochemical conclusions from the catastrophe model were:

a) the Moon must have approximately the same chemical composition as Earth’s mantle;

b) the Moon must have no core:

c) the Moon that was born from Earth must be younger than Earth;

d) as a result of the mega-impact, there should have been melting of the Moon and Earth. Thus, 
Earth and the Moon must have had oceans of liquid magma;

e) that the Moon is lacking in volatile elements is a consequence of the heating 
of Earth’s matter thrown into orbit upon the mega-impact.

In reality, according to the data of these geochemists, the picture is somewhat different: 

a) chemical composition of the Moon is notably different from that of Earth. In particular, 
the iron content in the Moon proved to be 1.5-2 times higher than in Earth’s mantle; 

b) the Moon has a significant core (1-3\% by mass, or 300-400 km in radius, which is about 20\% of 
the Moon’s radius of 1,738 km);

c) the Moon is older than Earth or, more correctly, the Moon’s core appeared earlier than Earth’s core;

d) the Moon was relatively cold and had been only partially flooded with magma. Geochemical data also refute 
the existence of an ocean of molten mantle on Earth. For example, the contemporary mantle of 
Earth is differentiated notably less than it would have been if an ancient molten magma ocean had existed;

e) content of volatile elements on the Moon does not support the mega-impact model. Their content cannot 
be derived from Earth’s mantle by heating \citep{Jones}.

But the criticism from space chemists was ignored, and the theory, none of whose space predictions came true, 
still remains almost universally accepted. Different variations of the giant impact theory seem to have 
provided solutions to the problem of increased iron content in the Moon’s rocks. According to these 
new theories, 80 percent of the Moon’s matter originated from the collision with 
Theia.  If  that  were  true,  however,  then  the  isotopic  composition  of  the  Moon 
would differ significantly from that of Earth (all space bodies have individual isotopic 
composition). Various groups of space chemists conducted analyses of the 
lunar minerals brought back from the Moon by the Apollo expeditions and found 
that the isotopic composition matched almost exactly with that of the Earth.

In order to explain this isotopic similarity, supporters of the mega-impact theory 
began to propose ever more catastrophic models, in which the mass of Theia 
increased from 10–20 percent to 30–45 percent of Earth’s mass \citep{Canup2}. 
This configuration better fitted the isotopic similarity, but raised a new problem of excessive 
angular momentum and did not address the existing problem of the absence of any 
signs of global melting on Earth.

In contrast, researchers who addressed the problem differently, by considering 
that the mass of Theia was actually less than ten percent that of Earth, solved the 
problem of the isotopic composition easily, because it would have been mostly 
matter from Earth rather than Theia that would have been thrown out to form the 
Moon in the event of a collision. Of course, if the mass of the impactor is less, the 
problem of the angular momentum deficit of the resulting Moon-Earth system 
arises. Scientists tried to overcome this problem by assuming that Earth’s rotation 
was very rapid before the mega-impact \citep{Stewart}. But does this not all suggest that too 
many  assumptions  are  being  made  in  order  to  save  the  mega-impact  theory?
Perhaps it would be better to wait until scientists overcome the intellectual barrier 
of the one-collision hypothesis and instead consider several collisions of not 
very large bodies which would not change the solution of the isotopic problem 
but would make it easier to solve the  angular momentum problem.
The solution to the problem of the formation of the asteroid belt is equally 
deplorable: over the past two hundred years, not a single model has been suggested 
that would clearly explain what happened to the majority of the mass from 
the asteroid belt.

Since 1978, there has been a chain of events that, sooner or later, will change 
the established paradigm of a catastrophic birth of our Moon. Charon, a moon 
of Pluto, the small, cold dwarf planet on the edge of the Solar System, was discovered 
in 1978. In terms of relative mass (12 percent of Pluto’s mass), Charon far 
outstrips the Moon (1.2 percent of Earth’s mass) and has thus deprived the Moon 
of its last unique characteristic (in terms of absolute mass, several satellites have 
been found around the giant planets which exceed that of the Moon).

In 1988, a group of astronomers led by V.V. Prokof’yeva-Mikhailovskaya from 
the Crimean Astrophysical Observatory began observations of the apparent magnitude  
of  asteroids.  The  characteristic  variability  recorded  for  the  asteroid  87 
Sylvia proved that the asteroid had a moon, and this was stated in a publication in 
1992 \citep{Prok}. The Crimean astronomers also found signs of duality with other asteroids, 
but nobody believed them. Asteroids were considered to be splinters; debris 
at the location of a planet that failed to form. The notion of satellites around 
asteroids was considered laughable.
The  laughter  stopped  instantly  when  the  interplanetary  station  Galileo  took 
photographs of the moon Dactyl around the asteroid Ida. As of April 2015, 270 
moons of asteroids and transneptunians (large bodies beyond Neptune; Pluto is 
now considered to be a transneptunian object) have been discovered. The asteroid 
87 Sylvia was found to have two moons and there are now several known triple 
asteroid systems. Pluto was found to have four more moons in addition to Charon, 
and became the first sextuple system of the asteroid type.
According to assessments by astronomers-observers, about 15 percent of asteroids 
have moons. This means that there is a rather regular, often realizable scenario 
for the formation of moons near small planets with a solid surface. 

How can this scenario be combined with the giant mega-impact theory for our own Moon? 
Perhaps, the Moon was formed in a similar way to other moons and is not actually 
unique?
Many asteroids, for example, Itokawa have a dumbbell form, or that of 
two agglutinated bodies. Apparently, in such cases, they are formed by the slow 
approach and attachment of the former moon and the primary body. Such nondestructive 
attachment between the asteroid and its moon is only possible if the 
orbital velocity of the moon is low, i.e., if the asteroid has a low mass. In the case 
of large asteroids, the merging of the moon with the central body would occur at 
a high velocity and end with the complete destruction of the moon. If the moon 
followed a circular orbit and the asteroid itself was large and quasi-spherical, the 
tangential collision of the moon would have left a long fossa (trench) on the surface 
of the primary body. Of note is that the large, fast-rotating asteroid Vesta 
(with an average diameter of 525 km) was found to have no moons, although two 
series of giant fossae were found on it – one in the equator area (Divalia Fossae 
Formation) and the other at $30\degr$N (Saturnalia Fossae Formation). The fossae reach 
300–400 km in length, 10–20 km in width, and up to 5 km in depth.

A moon should rotate around Vesta (near its surface) at a speed of 900 km/h. If 
its orbital rotation is prograde (coincides with the direction of the asteroid’s own 
rotation), the moon will move with a speed of about 600 km/h relative to Vesta’s 
surface. Assessments show that a moon with a mass a thousand times lower than 
that of Vesta, with the same density and a diameter of about 50 km, has enough 
kinetic energy before stopping (or full destruction) to make a canyon 5 km deep, 
10 km wide, and 1,000 km long even in strong granite, with a specific crushing 
energy of about $10^{9}$ erg/cm$^{3}$.
Such a moon could leave a fossa, in a material comparable with ice in terms of strength
($10^{8}$ erg/cm$^{3}$),
of 10,000 km long (or a shorter but  deeper  fossa).  The  two  series  of  grooves  
mean  that Vesta  swallowed  two 
moons, whose orbits were inclined to each other. Imagine a moon that flies above 
the asteroid’s surface at a speed of a turbojet airplane and crushes mountains several 
kilometers high – an impressive sight.

Another criteria is that of a more gentle attachment of a moon to an asteroid, 
assuming that the moon in this case has little energy and is incapable of digging a 
canyon (or crater) larger in volume than the moon itself. Thus, the conditions for 
forming  a  dumbbell  asteroid  would  include  a  restricted  orbital  velocity  of  the moon $V$:
$V^{2}<2E_v/\rho$, where $E_v$ is the specific volumetric energy of the asteroid’s destruction,  and 
 $\rho$ is  the  moon’s  density.  This  velocity  is  about  100 m/sec.
 Consequently, fast-rotating asteroids of up to several hundred thousand kilometers 
can have equatorial grooves, the footprints from the falls of their moons. For an 
asteroid with a diameter less than one hundred kilometers, merging with the moon 
must result in the formation of a dumbbell shape.
Vesta’s lack of moons has actively been discussed in specialist literature and 
popular journals. Journalist Jeff Hecht asked one of the authors of this appendix 
why Vesta has no moons. Having heard a hypothesis about Vesta engulfing its 
moons, he cited this hypothesis in an article on May 30, 2015, entitled: “Vesta has 
no moons—is it unlucky or did it eat them?” \citep{Hecht}

The process of moons and asteroids merging means that asteroids with moons 
could have been more numerous, possibly as much as half of the asteroid population.  
Perhaps  only  slowly-rotating  asteroids  and  planets  –  such  as Venus  and 
Mercury – did not have moons? Freshly-formed asteroids involved in strong collisions 
might also not have had sufficient time to acquire a moon.
In  1995,  Prokof’yeva-Mikhailovskaya,  Tarashchuk,  and  Gorkavyi  wrote  a 
review in Physics Physics-Uspekhi, in which, apart from describing the observational 
data of the Crimean astronomers, the dynamic stability of the orbits of the 
moons of asteroids that are located within the Hill sphere of their primary bodies 
was shown \citep{Prok2}.
 However, the reasons for the formation of relatively large moons 
around small asteroids with weak gravity remains unclear so far. The formation 
of our own Moon – comparatively large in relation to Earth – presented a similar  
problem,  but  the  mystery  of  small  moons  near  small  asteroids  was  most 
complicated.

Between 2004 and 2007, Gorkavyi proposed a single model for the formation 
of our Moon and those of the asteroids \citep{Gor1, Gor2}.
 According to this model, the Moon 
grew out of the regular circumplanetary cloud, whose mass multiplied due to a 
ballistic transfer of matter from Earth’s mantle (see Figure 1).
This transfer is similar to that used by the mega-impact theory, but in this model it occurs as several 
less catastrophic events rather than one mega-impact.

\begin{figure}
    \includegraphics[width=\columnwidth]{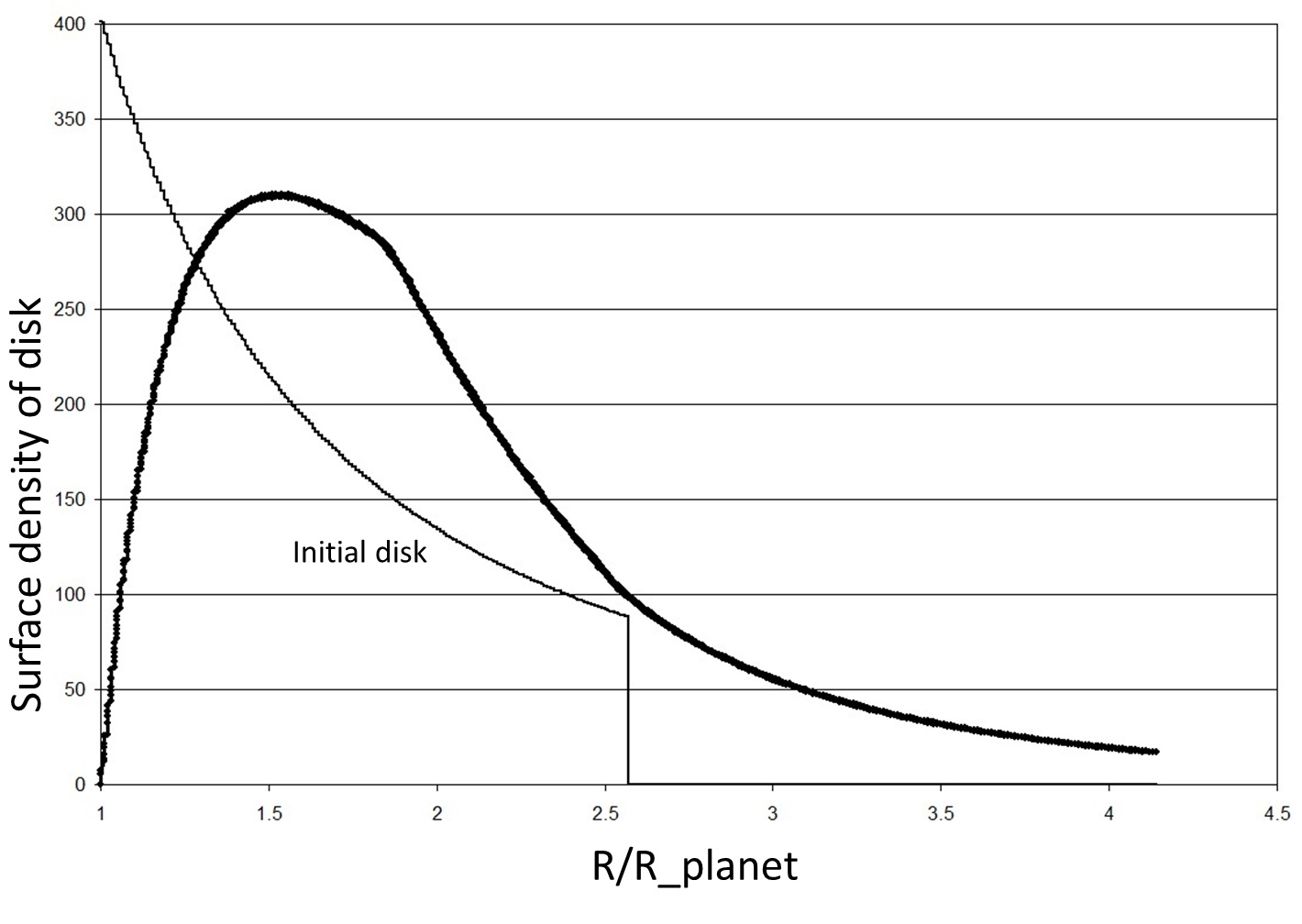}
    \caption{The density of the circumplanetary disk around the Earth grows because of 
the ballistic transfer of the ejecta from Earth’s surface.}
    \label{fig:fg1} 
\end{figure}

Taking into account the results obtained by the authors in other works focused 
on analytical and numerical research of the dust cloud dynamics, a general picture 
of the origin of the asteroid belt, the moons of asteroids, and our own Moon can 
be drawn \citep{Gor3, Gor4, Gor5}.

Asteroids lose their mass under micrometeorite bombardment. According to 
the experimental data of Japanese researchers, the asteroid Itokawa, which moves 
between Earth and Mars, loses many dozen centimeters of its surface layer over a 
million years \citep{Keis}.

Over 4.5 billion years, such loss would amount to a layer of 
2–3 km. Taking into account that the micrometeorite bombardment is more intensive 
in the zone of the asteroid belt and that the dust cloud density must have been 
much higher in the past, it can be concluded that asteroids can lose dozens and 
even hundreds of kilometers of their surface layers.

Where did all that mass, which turned into splinters and dust, go? If the asteroid 
belt was a closed system, nothing would happen; the dust would leave the surface 
of the asteroids and return without changing the general mass of the asteroid belt. 
However, in reality the asteroid belt is open to influence: it is lit by the Sun. The 
Sun’s rays create sufficient light pressure for the finest newly-formed dust grains 
of submicron size to be thrown out of the Solar System. Larger dust grains, of 1 
micron,  are  thrown  into  elliptical  orbits  immediately  after  their  formation  and 
approach Jupiter at their apocenters. Jupiter in turn treats them bluntly, throwing 
them in all directions at colossal speeds. Numerical calculations by the authors 
show that 50 percent of such dust grains are thrown out of the Solar System into 
interstellar space \citep{Gor4, Gor2}.

Jupiter, together with the solar pressure, works as a giant 
vacuum cleaner that removes fine dust from the asteroid belt. Larger dust particles, 
which do not reach Jupiter’s orbit and escape its ‘gravitational kick’, are 
affected differently and drift slowly to the Sun under the influence of the 
Poynting-Robertson effect. This effect makes a particle, which moves along the orbit in the 
media of solar wind quanta, lose its angular momentum (similarly, a man running 
during a vertical rain shower is hit in the face by raindrops due to his own speed) 
and crawl towards the Sun. This also decreases the asteroid belt mass.
The Sun and Jupiter are the ‘thieves’ that quietly and discreetly stole the main 
mass of the asteroid belt, preventing the material from forming a full-scale planet. 
Data from the asteroid Itokawa suggest that such a process is highly efficient.
Each  asteroid  was  gradually  reduced  over  cosmological  time,  sending  dust 
flows in all directions that, by its total mass, exceeded its current residue. From the 
point of view of Ilya Prigozhin’s concept of self-organization, this is a typical 
open system prone to the formation of structures. A reducing asteroid creates a 
space of opportunity around itself.
Moons are self-organizing structures that grow by feeding on the dust flying 
from asteroids. In fact, it is sufficient to create quite a small seed disk around an 
asteroid (the creation of such a disk from particles passing around planets was 
discussed in detail in the monograph Vityazev A.V.,  Pechernikova G.V., Safronov V.S. 
"Terrestrial planets: origin and early evolution", 1990)
and this begins to intercept the flying matter efficiently as it sorts: the 
retrograde  bodies  (in  terms  of  their  rotation  relative  to  the  disk  rotation)  are 
thrown to the planet and the prograde particles are captured by the disk, increasing 
its mass \citep{Gor2}.

Numerical calculations showed that even with isotropic bombardment of an 
asteroid by micrometeorites, a prograde disk, i.e. a disk rotating in the same direction 
as the asteroid itself, grows around it; such a disk has its peak density not far 
from the central body. As shown by research into the available observations, the 
main mass of the moons of asteroids is situated in the orbits equal to 3–4 radii of 
the central body (Figure 2).

\begin{figure}
    \includegraphics[width=\columnwidth]{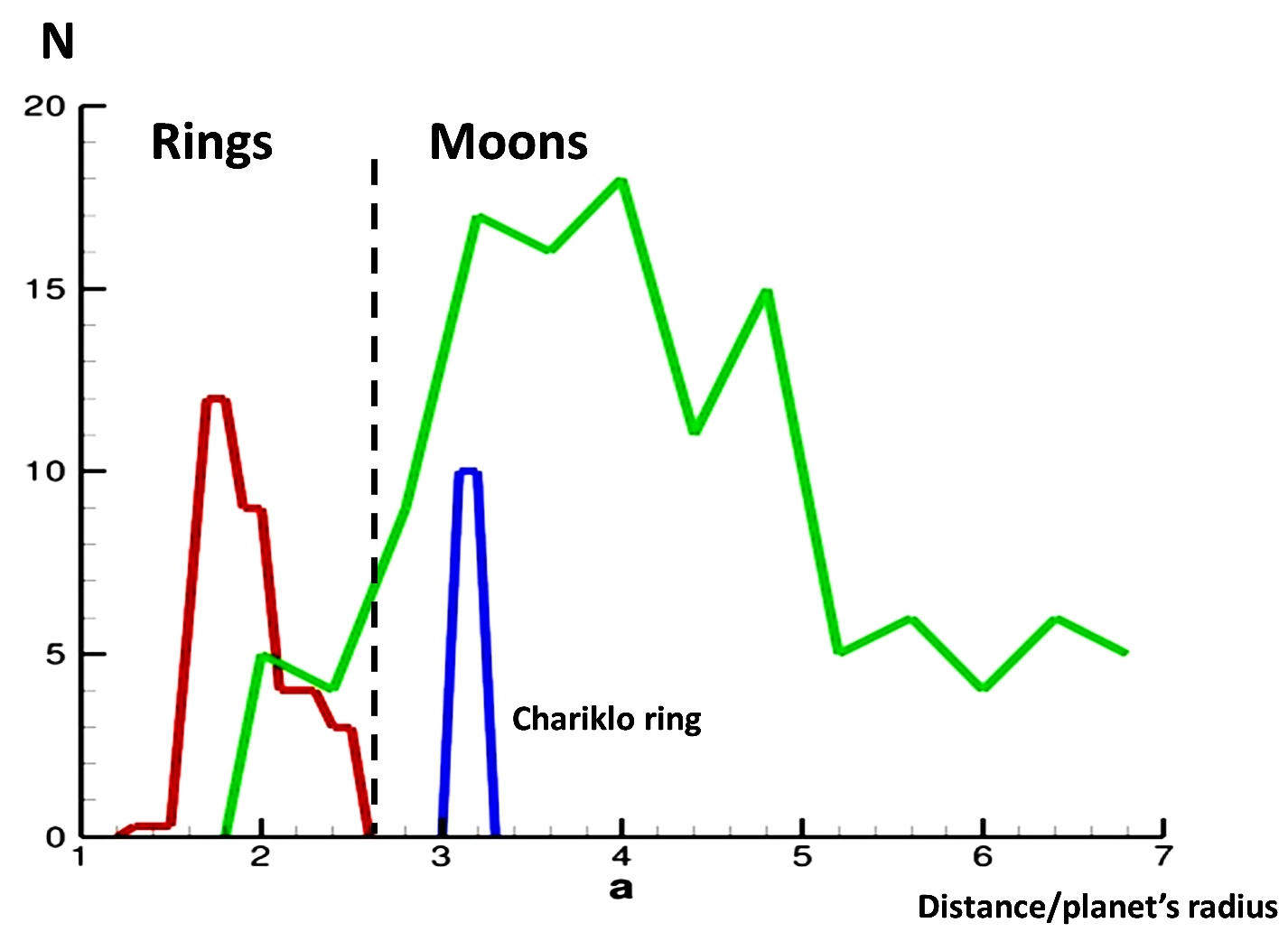}
    \caption{Distribution of planetary rings, moons of asteroids, and the Chariklo ring 
(the altitude of its peak is selected arbitrarily) by their distance from the planet. The dashed 
line means the outer borderline of planetary rings. It follows from this image that the 
Chariklo ring is a proto-moon disk rather than a planetary ring.}
    \label{fig:fg2} 
\end{figure}

The rings recently discovered near the Centaur-class asteroids Chariklo and 
Chiron were announced as planetary rings; however, in reality these are protomoon 
discs situated at a distance of exactly three radii from the asteroid, in contrast 
to planetary rings that are limited to a maximum radius of 2.6 (Figure 2).

How is the formation of our Moon different from the formation of the moons of 
asteroids? Actually, it does not differ in any principle respect. Micrometeorites 
and meteorites collide with an asteroid’s surface at a colossal speed and throw 
flows  of  friable  surface  matter  from  it. These  flows  have  much  slower  speeds 
(about 10 as shown by laboratory experiments \citep{Schultz}. In the case of the Moon, 
such micrometeorites and meteorites are replaced by larger bodies (planetesimals) 
from dozens of kilometers, up to 1,000 km. However, even 1,000-km planetesimals 
are hundreds of times lighter than Theia, meaning that the Moon appeared as 
a result of multiple moderate collisions, not a single catastrophic impact. The matter 
from Earth’s mantle was gradually thrown to orbit, to the proto-moon disk 
from which the Moon later formed, without the need for Earth to melt and without 
a unique mega-impact. The fact that the Moon formed quite near Earth and later 
moved to a distance equal to 60 Earth radii has been known for some time (the 
Moon is currently moving away from Earth at a speed of 4 cm per year).

One intriguing question is whether Earth had smaller, outer moons like those of 
Pluto. If so, then, the moving Moon must have engulfed its smaller brothers.
Thus, we have proposed a model that explains the three key problems of current 
planetology at once: the asteroid belt formation (or its mass loss); the appearance 
of multiple moons around asteroids (that were formed using the lost mass), 
and the birth of the Moon. This model does not require any unique events and, in 
the final analysis, better agrees with the geochemical data than the theory of a 
catastrophic mega-impact. This multi-impact model unites the most important 
and  significant  moments  of  the  mega-impact  theory  (throwing  Earth’s  mantle 
matter to space upon collision with a large body) and the accretion model (existence 
of a long-lived proto-satellite disk) and removes the difficulties of both 
concepts.
The  authors  have  learned  relatively  recently  that  the  theory  of  the  Moon’s 
origin as a result of a number of smaller impacts has also been developed independently 
by  the  Moscow  cosmogonic  group  of  Safronov-Ruskol-Vityazev-Pechernikova 
(see the history of this model in review \citet{Vit2}). This group also used 
the theory to develop the model of the Moon’s formation by accretion. According 
to the calculations of the Moscow specialists, the maximum size of bodies falling 
to  Earth  did  not  exceed  one  percent  of  Earth’s  mass,  meaning  they  were  of 
‘Lunar’, rather than ‘Martian’ size. This immediately renders the mega-impact 
model unrealistic.

An article by Gaftonyuk and Gorkavyi studied the patterns of the double asteroid 
database, which had already provided rather rich data \citep{Gor6}. It showed that the 
probability  of  a  moon’s  existence  grows  with  the  increased  rotation  speeds  of 
asteroids. Additionally, all the discovered moons rotate in the same direction as 
their central body and lie close to the plane of its equator. Thus, the moon systems 
of asteroids are regular and similar to those of the large planets.
In terms of relative mass, the Moon is in the middle of the distribution of moons 
of  solid  bodies. The  Moon  stopped  being  a  unique  body  in  the  planetological 
sense because many analogs for it have been discovered over time. Until recently, 
there were eight or nine planets in the Solar System and a lot of space debris. Now, 
we have millions of planets in our system, with their own origins, histories, and 
moons. Therefore, the models of moon formation near solid planetoids, including 
the Moon near Earth, must be unified to explain the formation of both the moons 
of asteroids and our own Moon. Any theory of the Moon’s formation that cannot 
explain the formation of the moons of asteroids is morally outdated, even if it has 
been suggested quite recently.
The  mega-impact  model  of  the  Moon’s  formation  is  also  connected  with 
the significant issue of the origin of Earth’s oceans. Where would the water have 
come from?

Tobi Owen, a Hawaiian astronomer and specialist in the origin of Earth’s hydrosphere,  
postulated  three  possible  sources  of  Earth’s  oceans  in  his  interview 
to Scientific American  of October 21, 1999: 1) Water that was contained in the 
asteroids (planetesimals) that agglutinated in the past to form the planet Earth. 
(This is, of course, a realistic source, because the primary bodies contained a lot 
of moisture.); 2) Water that was delivered to Earth during subsequent falls of 
asteroids and meteoroids such as carbonaceous chondrites from the asteroid belt; 3) 
Water from the comets that hit Earth infrequently \citep{Canup}.

In principle, asteroids and comets as sources of Earth’s water are realistic. The 
question is how much water has been delivered this way. Space chemists researched 
this issue and observed that if water had been delivered to Earth by meteorites, 
which contain a lot of the inert gas xenon, then Earth’s atmosphere would likely 
have contained ten times more xenon than it does. As the inert gas xenon does not 
engage in chemical reactions, its quantity has not changed since the time of Earth’s 
formation. Thus,  the  ‘meteoritic  hypothesis’  of  the  origin  of  Earth’s  oceans  is 
unlikely. Space chemists also criticized the ‘comet hypothesis’, because the deuterium/hydrogen  
ratio  in  cometary  ice  is  twice  as  high  as  in  Earth’s  oceans. 
Relatively  recent  measurements  taken  by  Rosetta  near  the  
Churyumov-Gerasimenko comet indicated that the deuterium/hydrogen ratio in that comet was 
three times higher than in Earth’s water, so any relationship between comets and 
Earth’s oceans is not confirmed \citep{Altwegg}.

 In their article ‘Oceans from Heaven’ in 
Scientific American  of March 2015, D. Jewitt and E. Young suggested that there is 
“no simple solution” to the choice between the comet hypothesis and the meteoritic 
hypothesis for the origin of oceans. However, why are researchers not happy 
with the first hypothesis, the simplest and most logical among the possibilities that 
were enumerated by Owen?
It is because the mega-impact model, which only creates the Moon with great 
difficulty,  also  deprives  Earth  of  its  oceans. As  Jewitt  and Young  noted:  “The 
energy of this global-scale impact would have swept away much of the atmosphere, 
flash-boiled any watery oceans and produced an ocean of magma hundreds of kilometers 
deep. Regardless of whether Earth formed wet or dry, the 
devastating blow of this Moon-forming impact must have cleansed our planet of 
nearly all its primordial water.”
Having  adopted  the  mega-impact  catastrophic  theory,  scientists  started  
racking their brains to devise sophisticated scenarios for returning the oceans to 
Earth.  There  is  no  need  to  rack  one’s  brains,  but  there  is  a  need  to  break 
stereotypes.
For any young scientists reading, catastrophism is a true sign of the scientific 
inadequacy of a theory. A theory based on the assumption of a rare event is 
inherently not good, because it has a low probability. What is worse is that a 
catastrophic event provides many free parameters for a model which on the one 
hand allows easy adaptation of a parametric model to the observed phenomenon, 
but on the other hand drastically decreases the scientific value of such a 
flexible model.
Catastrophic models have a low prediction value, because they are not applicable 
to other physical systems and cannot say anything about them. As a rule, 
catastrophic theories are debunked over time; if they have not been debunked so 
far, it simply means that not enough time has passed and that some of the proponents 
of the old model remain to defend it.

Let  us  consider  Jupiter’s  irregular  moons  and  the  theory  of  the  accidental 
capture by Neptune of its moon Triton, which has a retrograde orbit, as examples 
of the catastrophic formation model. According to the first model, the group of 
prograde (rotating in the same direction as the planet) and retrograde moons of 
Jupiter appeared due to the accidental collision of two large asteroids on head-on 
trajectories.
In 1993–95, the authors developed a numerical model to analyze the capture of 
small asteroids near a giant planet \citep{Gor7}.
 The model took into account the existence 
of a disc of prograde particles orbiting around the planet, and of collisions between 
particles of this disc and asteroids flying by. All the typical trajectories of passing 
asteroids were studied and hundreds of thousands of the splinter trajectories that 
formed from the collision of particles and asteroids were calculated, with different 
assumptions of their masses. The model was used for the systems of the three 
giant planets Jupiter, Saturn, and Neptune.
Surprisingly, it was found that all small asteroids are captured in quite specific 
zones near each planet, which are determined by the different geometry of the 
trajectories of the oncoming bodies. These zones are different for prograde and 
retrograde asteroids, and are where the real groups of retrograde and prograde 
irregular moons are located. 

Thus, the position of external moons that were considered 
to be irregular for a long time was found to be compliant with stringent 
celestial-mechanical regularities. The model explained the existence of Jupiter’s 
outer moons including the retrograde group Pasiphae, Saturn’s retrograde moon 
Phoebe, and Neptune’s big retrograde moon Triton. The diversity of the moon 
systems of Jupiter, Saturn, and Neptune proved to be connected to changes in only 
one significant parameter of the model – the evolution duration. Increasing evolution 
durations resulted in the formation of ever more massive retrograde moons. 
Therefore, Saturn has more such moons than Jupiter, and Neptune’s retrograde 
moon bodies dominated over prograde bodies and formed the giant retrograde 
Triton.
The article made a conclusion, based on the model for Saturn, that the planet 
had an undiscovered group of outer retrograde moons at twice the distance of the 
orbit radius of the retrograde Phoebe (13 million km), similar to Jupiter’s outer 
group Pasiphae (Figure 3).

\begin{figure}
    \includegraphics[width=\columnwidth]{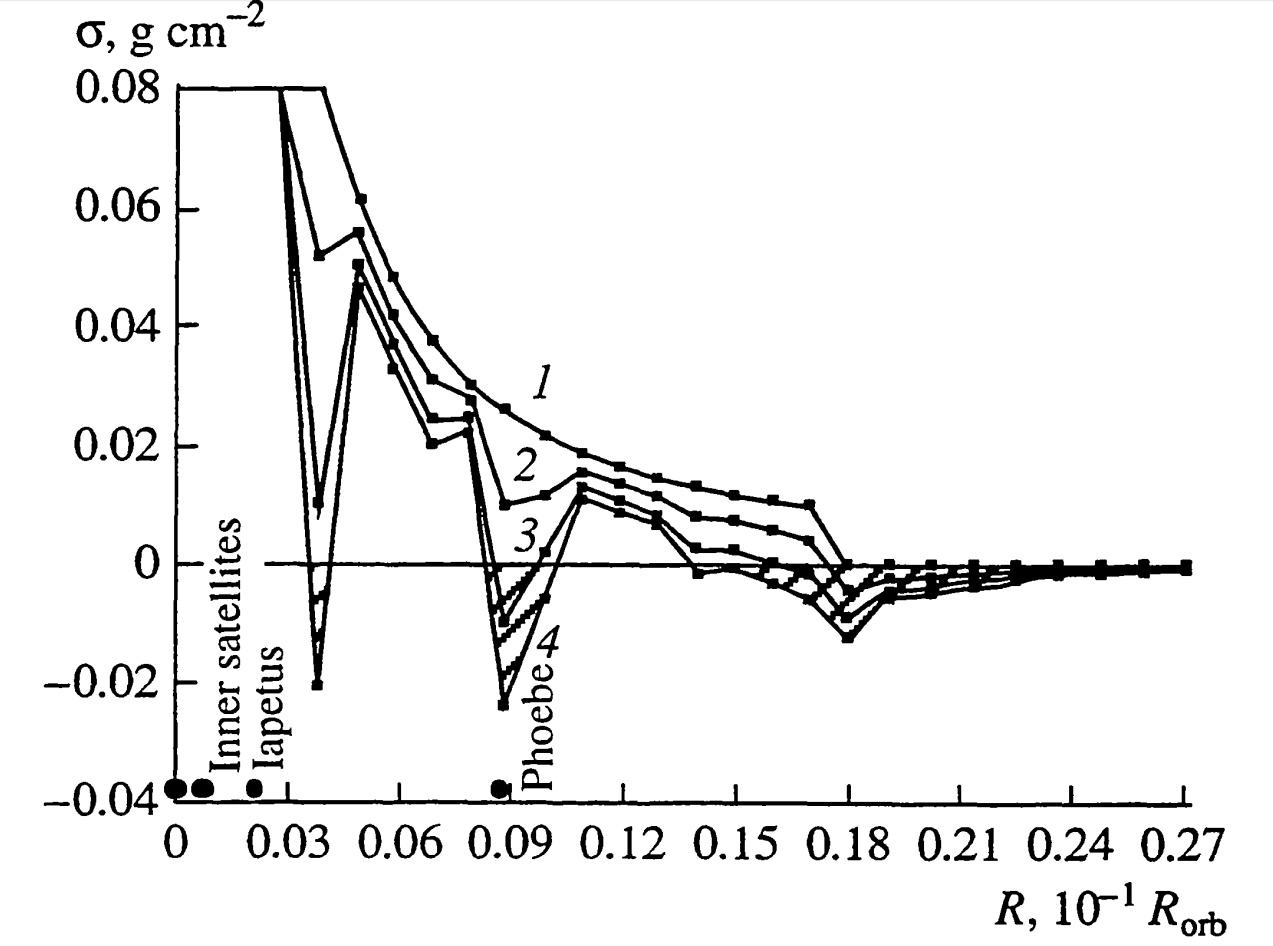}
    \caption{The original density profile (1) of the prograde near-planet disk evolves 
due to the capture of retrograde asteroids and their fragments (profiles 2–4) up to the 
zone where retrograde moons dominate (shaded zones below the axis). One such zone 
happens  to  be  around  the  orbit  of  Phoebe,  while  the  other,  outer  zone  contained  no 
known moons in 1995. A whole group of retrograde moons of Saturn were discovered in 
that zone some years later.}
    \label{fig:fg3} 
\end{figure}

 The prediction of the existence of this outermost group 
of retrograde moons was confirmed several years later: 25 retrograde moons of 
Saturn were discovered over 2000–2007, at a distance of 18–24 million km.
In 2001, the authors made an additional prediction, based on their 1995 calculations, 
that Nereid, the outermost moon of Neptune at that time, was the largest of 
the prograde moons in a group of outer moons, which would consist of a mixture 
of moons with prograde and retrograde orbits and with the retrograde moons dominant
\citep{Gor8}.
 This prediction was confirmed with the discovery of two prograde and 
three retrograde moons of Neptune beyond Nereid’s orbit in 2002–2003.
Thus, developments in science have systematically expelled catastrophic models, 
replacing them with theories that are based on regular rather than accidental 
phenomena.
The concept of the formation of the Moon, double asteroids, and the asteroid 
belt  itself  through  multiple  systematic  space  collisions  contains  a  lot  of 
little- studied and unclear phenomena, which is exactly why it presents a great 
opportunity  for  young  scientists  and  postgraduates,  including  those  from 
Chelyabinsk State University, to apply their abilities.











\bsp	
\label{lastpage}
\end{document}